\documentclass[US Letter,9pt]{article}
\usepackage{spconf,amsmath,graphicx}

\usepackage{graphicx}
\usepackage{bm}
\usepackage{amsfonts}
\usepackage{amsmath}
\usepackage{amssymb}
\usepackage{times}
\usepackage{cite}
\usepackage{subfigure}
\usepackage{latexsym,bm,amsmath,amssymb} 
\usepackage{CJK}
\usepackage{hhline}

\usepackage{multirow} 
\usepackage{amsbsy}
\usepackage{xcolor}
\usepackage[linesnumbered,ruled,vlined]{algorithm2e}
\usepackage{epstopdf}
\usepackage[font=small,labelfont=bf]{caption}
\usepackage{url}

\usepackage{geometry}
\title{\vspace{-5mm}Safeguarding UAV Networks Through Integrated Sensing, Jamming, and Communications\vspace{-5mm}}
\name{\centering Zhiqiang Wei$^{\star}$, Fan Liu$\dag$, Derrick Wing Kwan Ng$^{\ddagger}$, and Robert Schober$^{\star}$\vspace{-5mm}}
\address{$^{\star}$ Institute for Digital Communications (IDC),  Friedrich-Alexander\\ University Erlangen-Nuremberg, Germany
	(email: zhiqiang.wei; robert.schober@fau.de)\\
	$\dag$ Department of Electrical and Electronic Engineering \\ Southern University of Science and Technology, China (email: liuf6@sustech.edu.cn) \\ 
	$^{\ddagger}$ School of Electrical Engineering and Telecommunications \\
	University of New South Wales, Australia (email: w.k.ng@unsw.edu.au)\\
(\textit{Invited Paper})\vspace{-7mm}}
\begin{document}
\maketitle
\begin{abstract}
	\vspace{-2mm}
This paper proposes an integrated sensing, jamming, and communications (ISJC) framework for securing unmanned aerial vehicle (UAV)-enabled wireless networks.
The proposed framework advocates the dual use of artificial noise transmitted by an information UAV for simultaneous jamming and sensing of an eavesdropping UAV.
Based on the information sensed in the previous time slot, an optimization problem for online resource allocation design is formulated to maximize the number of securely served users in the current time slot, while taking into account a tracking performance constraint and quality-of-service (QoS) requirements regarding the leakage information rate to the eavesdropper and the downlink data rate to the legitimate users.
A channel correlation-based algorithm is proposed to obtain a suboptimal solution for the design problem.
Simulation results demonstrate the security benefits of integrating sensing into UAV communication systems.
\vspace{-2mm}
\end{abstract}
\begin{keywords}
UAV, physical layer security, extended Kalman filter, resource allocation.
\end{keywords}

\vspace{-4mm}
\section{Introduction}
\vspace{-4mm}
\label{sec:intro}
Compared with traditional terrestrial networks, UAV-enabled wireless networks \cite{wu20205g} have the potential to cover a larger area with a higher data rate, thanks to their high flexibility and mobility as well as the line-of-sight (LoS) dominated propagation channels to the ground terminals.
It is expected that UAV-enabled wireless networks will play a key role in supplementing future cellular networks by providing communication services to rural and hot spot areas \cite{ZhiqiangUAV}.
However, UAV communications are highly susceptible to eavesdropping as the associated leakage channels are also LoS dominated \cite{QingqingPLSUAV}.

Therefore, the exchange of confidential information in UAV wireless networks has to be safeguarded.
Numerous works studied secure resource allocation design for UAV-enabled wireless communication systems assuming the availability of perfect channel state information (CSI) of the eavesdropper channels \cite{QianPRLUAV,GuangchiUAV,XiaoboUAV}.
For the case of imperfect CSI, the authors of\cite{CaiEEUAV,MiaoUAV} proposed robust and secure resource allocation schemes to enhance the physical layer security (PLS) of UAV communications.
However, all these works \cite{QianPRLUAV,GuangchiUAV,XiaoboUAV,CaiEEUAV,MiaoUAV} assumed static eavesdroppers and did not develop specific sensing schemes for the eavesdropper channels.
In practice, if the eavesdropper has high-maneuverability, such as an eavesdropping UAV (E-UAV), guaranteeing communication security is very challenging and it is necessary to integrate leakage channel sensing/tracking into the system design.
The general concept of sensing-aided PLS has been discussed in the recent article \cite{wei2021towards}, but a corresponding practical scheme has not been reported.
The authors in \cite{su2021secure} studied PLS in dual-functional radar-communication (DFRC) systems, where the multi-user interference is designed to be constructive at the legitimate users, while disrupting the eavesdropper.
A jamming-based PLS enhancing scheme has been proposed for DFRC systems in \cite{NanchiTWC}. 
However, this scheme does not sense the eavesdropper channel.

In this paper, we propose an integrated sensing, jamming, and communications (ISJC) framework to guarantee PLS in UAV-enabled wireless networks.
In particular, an information UAV (I-UAV) transmits confidential information to legitimate ground users (GUs) and jams the E-UAV with artificial noises (AN) to facilitate secure communications.
At the same time, the I-UAV tracks the location and velocity of the E-UAV by reusing the reflected AN.
Based on the sensing information obtained in the previous time slot, a channel correlation-based online resource allocation algorithm is developed to determine the user scheduling and precoding policy for the current time slot.
Our simulation results demonstrate the benefits of integrating sensing, jamming, and communications for safeguard UAV networks.

\vspace{-4mm}
\section{System Model}
\vspace{-4mm}
\label{sec:SysModel}

\begin{figure}[t]
	\center{\includegraphics[width=2.8in]{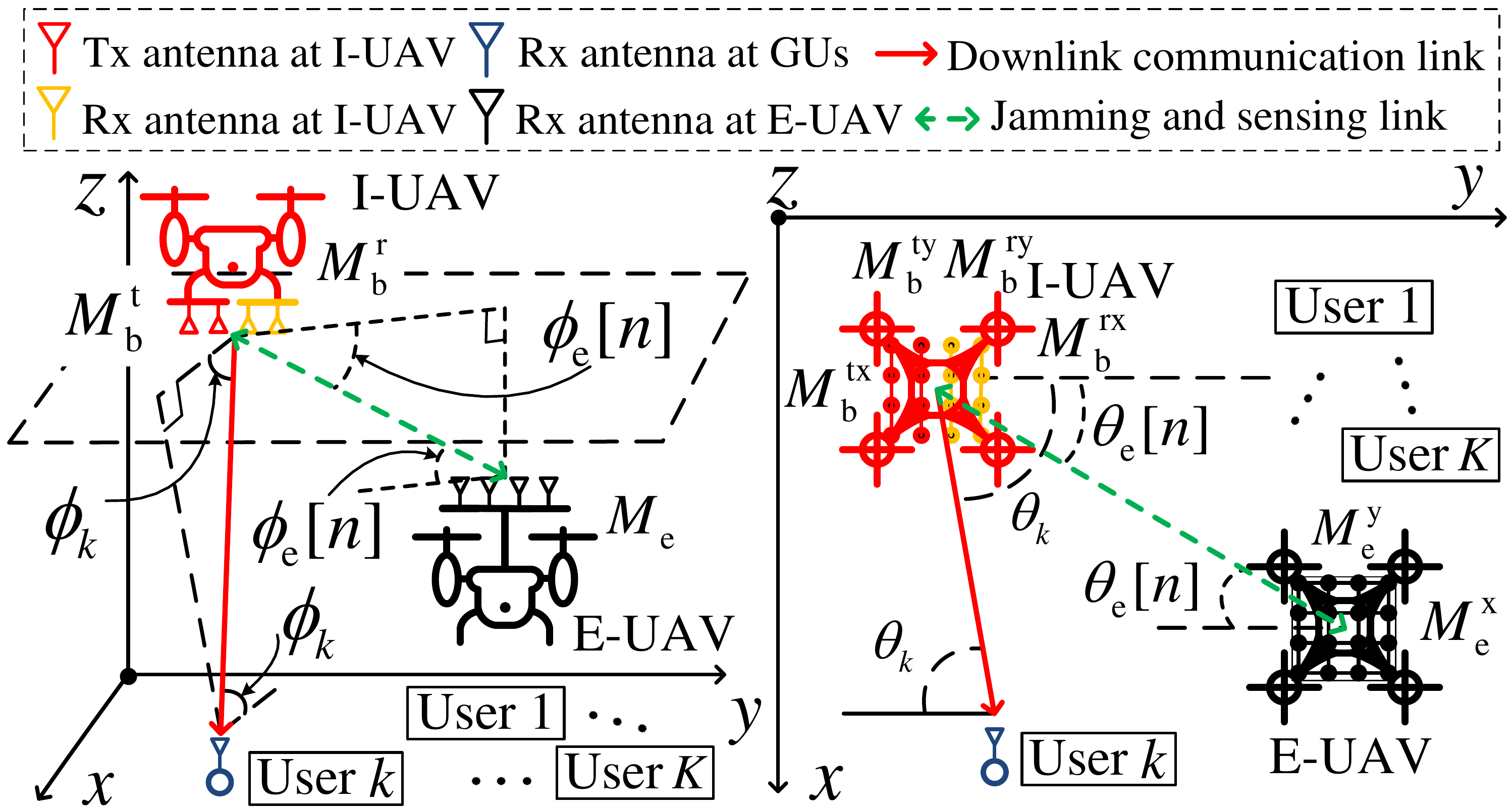}}\vspace{-3mm}
	\caption{A downlink UAV communication system serving $K$ GUs in the presence of an E-UAV.}\vspace{-8mm}
	\label{fig:SystemModel}
\end{figure}

\begin{figure}[t]
	\center{\includegraphics[width=3.2 in]{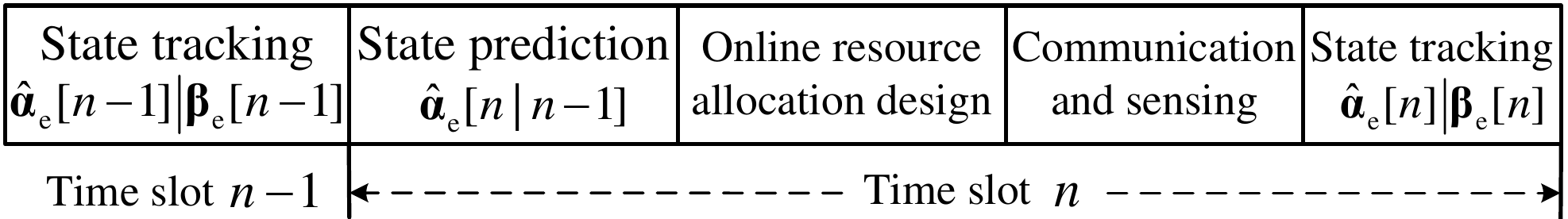}}\vspace{-3.5mm}
	\caption{The time slot structure of the proposed ISJC scheme.}\vspace{-7mm}
	\label{fig:TrackingModel}
\end{figure}

We consider a downlink UAV communication system where a hovering I-UAV\footnote{Mobile I-UAVs can further enhance PLS and will be considered in our future work.} serves as a base station broadcasting $K$ independent confidential data streams to $K$ legitimate single-antenna GUs in the presence of a flying E-UAV, cf. Fig. \ref{fig:SystemModel}.
%
%
The I-UAV is equipped with two uniform planar arrays (UPAs) for 3-dimensional (3D) transmit (Tx) and receive (Rx) beamforming, respectively, comprising
$M^{\mathrm{t}}_{\rm{b}}$ ($M^{\mathrm{tx}}_{\rm{b}}$ rows and $M^{\mathrm{ty}}_{\rm{b}}$ columns) and $M^{\mathrm{r}}_{\rm{b}}$ ($M^{\mathrm{rx}}_{\rm{b}}$ rows and $M^{\mathrm{ry}}_{\rm{b}}$ columns) antennas, respectively.
The E-UAV is equipped with a Rx UPA, comprising $M_{\rm{e}}$ ($M^{\rm{x}}_{\rm{e}}$ rows and $M^{\rm{y}}_{\rm{e}}$ columns) antennas.
The total service time $T$ is divided into $N$ equal-length time slots with a slot duration of $\delta$, i.e., $T \hspace{-1mm}=\hspace{-1mm} N\delta$.
%
%
%
The locations of the I-UAV and GUs are denoted as $\mathbf{q}_{\mathrm{b}} \hspace{-1mm}=\hspace{-1mm} \left[x_{\mathrm{b}},\hspace{-0.25mm}y_{\mathrm{b}},\hspace{-0.25mm}z_{\mathrm{b}}\right]^{\mathrm{T}}$ and $\mathbf{q}_k \hspace{-1mm}=\hspace{-1mm} \left[x_k,\hspace{-0.25mm}y_k,\hspace{-0.25mm}0\right]^{\mathrm{T}}$, $\forall k$, respectively, where $[ \cdot ]^{\mathrm{T}}$ is the transpose operation.
%
The E-UAV flies along a given trajectory, $\mathbf{q}_{\mathrm{e}}[\hspace{-0.25mm}n\hspace{-0.25mm}] \hspace{-1mm}=\hspace{-1mm} \left[x_{\mathrm{e}}\hspace{-0.25mm}[\hspace{-0.25mm}n\hspace{-0.25mm}],\hspace{-0.25mm}y_{\mathrm{e}}\hspace{-0.25mm}[\hspace{-0.25mm}n\hspace{-0.25mm}],\hspace{-0.25mm}z_{\mathrm{e}}\hspace{-0.25mm}[\hspace{-0.25mm}n\hspace{-0.25mm}]\right]^{\mathrm{T}}$,
designed for intercepting the legitimate information transmission with velocity $\dot{\mathbf{q}}_{\mathrm{e}}\hspace{-0.25mm}[\hspace{-0.25mm}n\hspace{-0.25mm}] \hspace{-1mm}=\hspace{-1mm} [\dot{x}_{\mathrm{e}}\hspace{-0.25mm}[\hspace{-0.25mm}n\hspace{-0.25mm}],\hspace{-0.25mm}\dot{y}_{\mathrm{e}}\hspace{-0.25mm}[\hspace{-0.25mm}n\hspace{-0.25mm}],\hspace{-0.25mm}\dot{z}_{\mathrm{e}}\hspace{-0.25mm}[\hspace{-0.25mm}n\hspace{-0.25mm}]]^{\mathrm{T}}$, $n\hspace{-1mm} =\hspace{-1mm} \left\{1,\hspace{-0.25mm}\ldots,\hspace{-0.25mm}N\right\}$.
The state of the E-UAV, $\boldsymbol{\alpha}_{\mathrm{e}}\hspace{-0.25mm} [n] \hspace{-1mm} = \hspace{-1mm} [\mathbf{q}^{\mathrm{T}}_{\mathrm{e}}\hspace{-0.25mm}[\hspace{-0.25mm}n\hspace{-0.25mm}],\hspace{-0.25mm} \dot{\mathbf{q}}^{\mathrm{T}}_{\mathrm{e}}\hspace{-0.25mm}[\hspace{-0.25mm}n\hspace{-0.25mm}]]^{\mathrm{T}}$, is unknown to the I-UAV.
In contrast, the locations of both the I-UAV and GUs are known to the E-UAV.
The time slot structure of the proposed ISJC scheme is illustrated in Fig. \ref{fig:TrackingModel}, where $\boldsymbol{\beta}_{\mathrm{e}} \hspace{-0.25mm}[\hspace{-0.25mm}n\hspace{-0.25mm}]$ represents all observable parameters at E-UAV.
At the beginning of time slot $n$, I-UAV first predicts the state of E-UAV based on the state estimates in time slot $n\hspace{-1mm}-\hspace{-1mm}1$, i.e., $\hat{\boldsymbol{\alpha}}_{\mathrm{e}}\hspace{-0.25mm}[\hspace{-0.25mm}n|n\hspace{-1mm}-\hspace{-1mm}1]$.
%
%
Based on the predicted state, I-UAV designs the resource allocation policy for time slot $n$ and broadcasts the information and the AN accordingly.
Exploiting the echoes from E-UAV, I-UAV takes the new measurement $\boldsymbol{\beta}_{\mathrm{e}} \hspace{-0.25mm}[\hspace{-0.25mm}n\hspace{-0.25mm}]$ and estimates the new state of E-UAV $\hat{\boldsymbol{\alpha}}_{\mathrm{e}} \hspace{-0.25mm}[\hspace{-0.25mm}n\hspace{-0.25mm}]$, which is used as the input of the predictor for time slot $n\hspace{-1mm}+\hspace{-1mm}1$.
Note that the considered system is inherently causal, where the resource allocation design in time slot $n$ is based on new measurements and new state estimates in time slot $n\hspace{-1mm}-\hspace{-1mm}1$.
%
%
%

The received signal at GU $k$ in time slot $n$ is given by
\vspace{-3.5mm}
\begin{align}\label{ReceivedSignalModel}
{y}_k(n,t) &= {\sum\nolimits_{k' = 1}^{K} \mathbf{h}^{\mathrm{H}}_{\mathrm{b},k}\mathbf{w}_{k'} [n] u_{k'}[n] s_{k'} (n,t)} \notag\\[-1mm]
&+ {\mathbf{h}^{\mathrm{H}}_{\mathrm{b},k}\mathbf{w}_{\mathrm{e}} [n] a_{\mathrm{e}} (n,t)} + {v}_k(n,t),
\end{align}
\vspace{-8mm}\par\noindent
where $t \hspace{-1mm}\in \hspace{-1mm}\left(\hspace{-0.25mm}0,\hspace{-0.25mm}\delta\hspace{-0.25mm}\right)$ is a time instant within time slot $n$, $s_k (n,t) \hspace{-1mm}\sim \hspace{-1mm}\mathcal{CN}\hspace{-1mm} \left(\hspace{-0.25mm}0,\hspace{-0.25mm}1\hspace{-0.25mm}\right)$ denotes the transmitted information signal for GU $k$, $a_{\mathrm{e}} (n,t) \hspace{-1mm}\sim \hspace{-1mm}\mathcal{CN}\hspace{-1mm} \left(\hspace{-0.25mm}0,\hspace{-0.25mm}1\hspace{-0.25mm}\right)$ is the AN, and ${v}_k(n,t) \hspace{-1mm}\sim\hspace{-1mm} \mathcal{CN}\hspace{-1mm}\left(\hspace{-0.25mm}0,\hspace{-0.25mm}\sigma_k^2\hspace{-0.25mm}\right)$ denotes the background noise at GU $k$ with power $\sigma_k^2$.
Here, $\mathcal{CN}\hspace{-0.5mm}(\hspace{-0.25mm}\mu,\hspace{-0.25mm}\sigma^2\hspace{-0.25mm})$ denotes the circularly symmetric complex Gaussian (CSCG) distribution with mean $\mu$ and variance $\sigma^2$.
$u_{k}[n] \hspace{-1mm}=\hspace{-1mm} 1$ indicates that GU $k$ is selected for communication in time slot $n$, otherwise, $u_{k}[n] \hspace{-1mm}=\hspace{-1mm} 0$.
Vectors $\mathbf{w}_k [n] \hspace{-1mm}\in\hspace{-1mm} \mathbb{C}^{M^{\mathrm{t}}_{\mathrm{b}} \times 1}$ and $\mathbf{w}_{\mathrm{e}} [n] \hspace{-1mm}\in\hspace{-1mm} \mathbb{C}^{M^{\mathrm{t}}_{\mathrm{b}} \times 1}$ denote the precoding vectors for GU $k$ and AN, respectively.
Vector $\mathbf{h}_{\mathrm{b},k} \hspace{-1mm}=\hspace{-1mm}\frac{\beta_0}{d_k} \mathbf{A}_{M_{\mathrm{b}}^{\mathrm{tx}},M_{\mathrm{b}}^{\mathrm{ty}}}\hspace{-0.5mm}\left(\hspace{-0.5mm}\theta_{k},\hspace{-0.5mm}\phi_{k}\hspace{-0.5mm}\right)$ is the channel vector between I-UAV and GU $k$, where $\mathbf{A}_{M^{\mathrm{x}}\hspace{-0.25mm},M^{\mathrm{y}}}\hspace{-0.5mm}\left(\hspace{-0.5mm}\theta, \hspace{-0.5mm}\phi\hspace{-0.5mm}\right) \hspace{-1mm}\in\hspace{-1mm} \mathbb{C}^{M^{\mathrm{x}}M^{\mathrm{y}} \hspace{-0.5mm}\times\hspace{-0.5mm} 1}$ is the steering vector of a UPA of size $M^{\mathrm{x}}\hspace{-1mm}\times \hspace{-1mm}M^{\mathrm{y}}$ \cite{van2004optimum}.
Constant $\beta_0$ represents the channel gain at a reference distance and $d_k \hspace{-0.75mm}=\hspace{-0.75mm} \|\mathbf{q}_{\mathrm{b}} \hspace{-1mm}-\hspace{-1mm} \mathbf{q}_{k}\|$ denotes the distance between I-UAV and GU $k$, where $\left\| \cdot \right\|$ denotes the $l_2$-norm of a vector.
Note that as this is the first work on ISJC in UAV networks, we assume that all links are LoS-dominated to facilitate the presentation.
%
%
%
Angles $\theta_{k}\hspace{-1mm}\in\hspace{-1mm} \left[\hspace{-0.25mm}-\pi,\hspace{-0.25mm}\pi\hspace{-0.25mm}\right]$ and $\phi_{k} \hspace{-1mm}\in\hspace{-1mm} \left[\hspace{-0.25mm}-\pi/2,\hspace{-0.25mm}0\hspace{-0.25mm}\right)$ denote the azimuth angle of departure (AOD) and the elevation AOD from I-UAV to GU $k$, respectively, cf. Fig. \ref{fig:SystemModel}.

As the precoding policy of I-UAV is unknown to E-UAV, we assume that E-UAV performs maximum ratio combining (MRC) to maximize its received signal power.
Assuming the azimuth and elevation angles-of-arrival (AoAs) are perfectly known at the E-UAV, the post-processed signal at E-UAV in time slot $n$ is
\vspace{-3mm}
\begin{align}\label{EavUAVSignal}
{y}_{\mathrm{e}}(n,t) &= {\sum\nolimits_{k = 1}^{K} \sqrt{M_{\mathrm{e}}} \mathbf{h}^{\mathrm{H}}_{\mathrm{be}}[n] \mathbf{w}_k [n] u_{k}[n] s_k (n,t)} \notag\\[-1mm]
& +\sqrt{M_{\mathrm{e}}}{\mathbf{h}^{\mathrm{H}}_{\mathrm{be}}[n]\mathbf{w}_{\mathrm{e}} [n] a_{\mathrm{e}} (n,t)} + {v}_{\mathrm{e}}(n,t),
\end{align}
\vspace{-7mm}\par\noindent
where $M_{\mathrm{e}}$ represents the MRC gain and ${v}_{\mathrm{e}}(n,t) \hspace{-1mm}\in\hspace{-1mm} \mathcal{CN}\left(0,\sigma_{\mathrm{e}}^2\right)$ is the noise at E-UAV.
Vector $\mathbf{h}_{\mathrm{be}}[n]\hspace{-1mm}=\hspace{-1mm} \frac{\beta_0}{d_{\mathrm{e}}[n]}\mathbf{A}_{{M^{\mathrm{tx}}_{\mathrm{b}}}\hspace{-0.25mm},{M^{\mathrm{ty}}_{\mathrm{b}}}}\hspace{-1mm}\left(\hspace{-0.5mm}\theta_{\mathrm{e}}[n],\hspace{-0.5mm} \phi_{\mathrm{e}}[n]\hspace{-0.5mm}\right)$ denotes the effective channel between I-UAV and E-UAV, where $d_{\mathrm{e}}[n] \hspace{-1mm}= \hspace{-1mm}\|\mathbf{q}_{\mathrm{b}} \hspace{-1mm}-\hspace{-1mm} \mathbf{q}_{{\mathrm{e}}}[n]\|$ is the corresponding distance and $\theta_{\mathrm{e}}[n]\hspace{-1mm}\in\hspace{-1mm} \left[\hspace{-0.25mm}-\pi,\hspace{-0.25mm}\pi\hspace{-0.25mm}\right]$ and  $\phi_{\mathrm{e}}[n]\hspace{-1mm}\in\hspace{-1mm} \left[\hspace{-0.25mm}-\pi/2,\hspace{-0.25mm}\pi/2\hspace{-0.25mm}\right]$ denote the azimuth and elevation AODs from I-UAV to E-UAV, respectively.
Due to the unknown location of E-UAV, $\theta_{\mathrm{e}}\hspace{-0.25mm}[\hspace{-0.25mm}n\hspace{-0.25mm}]$, $\phi_{\mathrm{e}}\hspace{-0.25mm}[\hspace{-0.25mm}n\hspace{-0.25mm}]$, $d_{\mathrm{e}}\hspace{-0.25mm}[\hspace{-0.25mm}n\hspace{-0.25mm}]$, and $\mathbf{h}_{\mathrm{be}}\hspace{-0.25mm}[\hspace{-0.25mm}n\hspace{-0.25mm}]$ are a prior unknown and have to be estimated by I-UAV.


Based on \eqref{ReceivedSignalModel}, the achievable data rate of GU $k$ is given by
\vspace{-3mm}
\begin{equation}\label{DownlinkRate}
{R_k}\hspace{-0.5mm}[\hspace{-0.25mm}n\hspace{-0.25mm}] \hspace{-1mm}= \hspace{-1mm}{\log _2}\Bigg(\hspace{-1mm}{1 \hspace{-1mm}+\hspace{-1mm} \frac{{{{u_{k}[n]\big| {{\bf{h}}_{{\rm{b}},k}^{\rm{H}}{{\bf{w}}_k}[n]} \big|}^2}}}{{\sum\limits_{k' \neq k}\hspace{-1mm} {{{u_{k'}\hspace{-0.5mm}[\hspace{-0.25mm}n\hspace{-0.25mm}]\hspace{-0.5mm}\big|\hspace{-0.25mm} {{\bf{h}}_{{\rm{b}},k}^{\rm{H}}\hspace{-0.5mm}{{\bf{w}}_{k'}}\hspace{-0.5mm}[\hspace{-0.25mm}n\hspace{-0.25mm}]} \hspace{-0.5mm}\big|}^2}}  \hspace{-1mm}+\hspace{-1mm} {{\big|\hspace{-0.25mm} {{\bf{h}}_{{\rm{b}},k}^{\rm{H}}\hspace{-0.5mm}\hspace{-0.25mm}{{\bf{w}}_{\rm{e}}}\hspace{-0.5mm}[\hspace{-0.25mm}n\hspace{-0.25mm}]} \hspace{-0.5mm}\big|}^2} \hspace{-1mm}+\hspace{-1mm} {{\sigma _k^2}} }}} \hspace{-0.5mm}\Bigg)\hspace{-0.5mm}.\vspace{-3.5mm}
\end{equation}
When E-UAV intercepts the information of GU $k$, as a worst case, we assume it can mitigate the inter-user interference of the other GUs and is only impaired by the AN.
Thus, the leakage information rate associated with GU $k$ in time slot $n$ is given by
\vspace{-3.25mm}
\begin{equation}\label{EavesdroppingUAVRate}
{R^{k}_{\rm{e}}}[n] = {\log _2}\Bigg( {1 + \frac{{{{u_{k}[n]M_{\rm{e}}\big| {{\bf{h}}_{{\rm{be}}}^{\rm{H}}[n]{{\bf{w}}_k}[n]} \big|}^2}}}{{{{M_{\rm{e}}\big| {{\bf{h}}_{{\rm{be}}}^{\rm{H}}[n]{{\bf{w}}_{\rm{e}}}[n]} \big|}^2} + {\sigma _{\rm{e}}^2}}}} \Bigg).\vspace{-3mm}
\end{equation}

In practice, the signal transmitted by I-UAV is partially received by the Rx antennas of E-UAV and is partially reflected by its body.
The I-UAV is assumed to operate in the full-duplex mode, which allows it to transmit and receive signals simultaneously, while the received signal may suffer residual self-interference\cite{SunYanNOMA}.
The corresponding echo received at I-UAV in time slot $n$ is given by
\vspace{-4.5mm}
\begin{align}\label{EchoesUAV}
\mathbf{r}_{\mathrm{b}}\hspace{-0.25mm}(\hspace{-0.25mm}n,\hspace{-0.25mm}t\hspace{-0.25mm})&= { e^{j2\pi\nu_{\mathrm{e}}\hspace{-0.25mm}[n]t}\mathbf{H}^{\mathrm{r}}_{\mathrm{be}}[n] \hspace{-0.75mm}\sum\nolimits_{k = 1}^{K} \hspace{-1mm} \mathbf{w}_k \hspace{-0.25mm}[\hspace{-0.25mm}n\hspace{-0.25mm}] \hspace{-0.25mm}u_k\hspace{-0.25mm}[\hspace{-0.25mm}n\hspace{-0.25mm}]s_k\hspace{-0.25mm} (\hspace{-0.25mm}n,\hspace{-0.25mm}t\hspace{-1mm}-\hspace{-1mm}\tau_{\mathrm{e}}\hspace{-0.25mm}[\hspace{-0.25mm}n\hspace{-0.25mm}]\hspace{-0.25mm})}  \notag\\[-1mm]
&+ { e^{j2\pi\nu_{\mathrm{e}}[\hspace{-0.25mm}n\hspace{-0.25mm}]t}\mathbf{H}^{\mathrm{r}}_{\mathrm{be}}[\hspace{-0.25mm}n\hspace{-0.25mm}] \mathbf{w}_{\mathrm{e}} [\hspace{-0.25mm}n\hspace{-0.25mm}] a_{\mathrm{e}} (n,t\hspace{-1mm}-\hspace{-1mm}\tau_{\mathrm{e}}[\hspace{-0.25mm}n\hspace{-0.25mm}])} \hspace{-0.5mm}+\hspace{-0.5mm} \mathbf{v}_{\mathrm{b}}\hspace{-0.25mm}(\hspace{-0.25mm}n,\hspace{-0.25mm}t\hspace{-0.25mm}),
\end{align}
\vspace{-7mm}\par\noindent
where the round-trip channel matrix is given by
\vspace{-3.25mm}
\begin{equation}\label{eqn:RoundTripChannel}
\mathbf{H}^{\mathrm{r}}_{\mathrm{be}}[\hspace{-0.25mm}n\hspace{-0.25mm}] \hspace{-1mm}=\hspace{-1mm}\frac{\epsilon_{\mathrm{e}}\hspace{-0.25mm}[\hspace{-0.25mm}n\hspace{-0.25mm}]\beta_0}{2d_{\mathrm{e}}\hspace{-0.25mm}[\hspace{-0.25mm}n\hspace{-0.25mm}]} \hspace{-0.25mm} \mathbf{A}_{M_{\mathrm{b}}^{\mathrm{rx}}\hspace{-0.25mm},M_{\mathrm{b}}^{\mathrm{ry}}}\hspace{-1mm}\left(\hspace{-0.5mm}\theta_{\mathrm{e}}\hspace{-0.5mm}[\hspace{-0.25mm}n\hspace{-0.25mm}],\hspace{-0.25mm} \phi_{\mathrm{e}}\hspace{-0.5mm}[\hspace{-0.25mm}n\hspace{-0.25mm}]\hspace{-0.5mm}\right)\hspace{-1mm}\mathbf{A}^{\mathrm{H}}_{{M^{\mathrm{tx}}_{\mathrm{b}}}\hspace{-0.25mm},{M^{\mathrm{ty}}_{\mathrm{b}}}}\hspace{-1mm}\left(\hspace{-0.5mm}\theta_{\mathrm{e}}\hspace{-0.5mm}[\hspace{-0.25mm}n\hspace{-0.25mm}],\hspace{-0.25mm} \phi_{\mathrm{e}}\hspace{-0.5mm}[\hspace{-0.25mm}n\hspace{-0.25mm}]\hspace{-0.5mm}\right)\hspace{-1mm}.\vspace{-3mm}
\end{equation}
Here, variables $\tau_{\mathrm{e}}\hspace{-0.25mm}[\hspace{-0.25mm}n\hspace{-0.25mm}]$ and $\nu_{\mathrm{e}}\hspace{-0.25mm}[\hspace{-0.25mm}n\hspace{-0.25mm}]$ denote the round-trip time delay and Doppler shifts, respectively, $\epsilon_{\mathrm{e}}\hspace{-0.25mm}[\hspace{-0.25mm}n\hspace{-0.25mm}]\hspace{-1mm} =\hspace{-1mm} \sqrt{\hspace{-1mm}\frac{\vartheta_{\mathrm{e}}}{4\pi d^2_{\mathrm{e}}[\hspace{-0.25mm}n\hspace{-0.25mm}]}}$ denotes the reflection coefficient of E-UAV in time slot $n$, and $\vartheta_{\mathrm{e}}$ is the radar
cross-section of E-UAV \cite{skolnik1962introduction}.
Vector $\mathbf{v}_{\mathrm{b}}\hspace{-0.25mm}(\hspace{-0.25mm}n,\hspace{-0.25mm}t\hspace{-0.25mm}) \hspace{-1mm}\in\hspace{-1mm} \mathcal{CN}\hspace{-1mm} \left(\hspace{-0.25mm}\mathbf{0},\hspace{-0.25mm}\sigma_{\mathrm{b}}^2\mathbf{I}_{M^{\mathrm{r}}_{\mathrm{b}}}\hspace{-0.25mm}\right)$ captures both the background noise and the residual self-interference \cite{SunYanNOMA} at I-UAV, where $\mathbf{I}_{M^{\mathrm{r}}_{\mathrm{b}}}$ denotes an $M^{\mathrm{r}}_{\mathrm{b}} \times M^{\mathrm{r}}_{\mathrm{b}}$ identical matrix. 
Note that clutter, such as the reflected signals of the GUs and the ground itself, is omitted here as it can be substantially suppressed by clutter suppression techniques \cite{skolnik1962introduction} owing to its distinctive reflection angles and Doppler frequencies compared to the echoes from E-UAV \cite{skolnik1962introduction}.
Besides, we assume that the AOA is identical to the corresponding AOD at I-UAV in \eqref{eqn:RoundTripChannel}, which is reasonable when assuming a point target model \cite{FanLiuISAC} and reciprocal propagation.

%

\vspace{-4mm}
\section{E-UAV Tracking}
\label{subsec:Tracking Model}

\vspace{-4mm}
\subsection{Estimation Model of E-UAV}
\vspace{-3mm}
Based on the echoes in \eqref{EchoesUAV}, different estimation methods can be used to estimate $\tau_{\mathrm{e}}\hspace{-0.25mm}[\hspace{-0.25mm}n\hspace{-0.25mm}]$, $\nu_{\mathrm{e}}\hspace{-0.25mm}[\hspace{-0.25mm}n\hspace{-0.25mm}]$, $\phi_{\mathrm{e}}\hspace{-0.25mm}[\hspace{-0.25mm}n\hspace{-0.25mm}]$, and $\theta_{\mathrm{e}}\hspace{-0.25mm}[\hspace{-0.25mm}n\hspace{-0.25mm}]$ \cite{kay1993fundamentals}.
One possible approach to estimate these parameters is the matched filter (MF) principle by exploiting the AN \cite{kay1993fundamentals}:
\vspace{-3.2mm}
\begin{align}
&\left\{\hspace{-0.5mm} {{\hat \tau _{\rm{e}}}[n],{\hat\nu _{\rm{e}}}[n]}\hspace{-0.5mm},\hat \theta_{\mathrm{e}}[n],\hat\phi_{\mathrm{e}}[n] \right\}\hspace{-1mm} \label{TFSyn}\\[-2mm]
& = \arg \mathop {\max }\limits_{\tau ,\nu, \theta, \phi} \hspace{-0.5mm}\Big|\hspace{-0.5mm} \frac{1}{\delta}\hspace{-0.5mm}{\int_0^\delta\hspace{-2mm} \mathbf{A}^{\mathrm{H}}_{M_{\mathrm{b}}^{\mathrm{rx}},M_{\mathrm{b}}^{\mathrm{ry}}}\hspace{-0.5mm}\left(\hspace{-0.5mm}\theta,\hspace{-0.5mm}\phi\hspace{-0.5mm}\right)\hspace{-0.5mm} {{{\bf{r}}_{\rm{b}}}\hspace{-0.5mm}(\hspace{-0.5mm}n,\hspace{-0.5mm}t\hspace{-0.5mm})a_{\rm{e}}^ *\hspace{-0.5mm} (\hspace{-0.5mm}n,\hspace{-0.5mm}t \hspace{-0.5mm}- \hspace{-0.5mm}\tau \hspace{-0.5mm}){e^{ - j2\pi \nu t}}dt} } \Big|.\notag
\end{align}
\vspace{-5mm}\par\noindent
Note that we exploit the AN rather than the user signals for sensing as using user signals would require the I-UAV to beamform the information-bearing signals towards the E-UAV, which increases the risk of information leakage.
Instead, using AN for both sensing and jamming is a win-win strategy for secrecy applications.
Analyzing the estimation variances associated with \eqref{TFSyn} is a challenging task.
According to \cite{FanLiuISAC,kay1993fundamentals}, assuming the independence of the AN and the user signals, i.e., $\frac{1}{\delta}\hspace{-1mm}\int_0^\delta\hspace{-0.5mm}  s_k\hspace{-0.25mm}(\hspace{-0.25mm}n,\hspace{-0.25mm}t\hspace{-0.25mm}) a^*_{\mathrm{e}}\hspace{-0.25mm}(\hspace{-0.25mm}n,\hspace{-0.25mm}t\hspace{-0.25mm}) dt \hspace{-0.5mm} \approx \hspace{-0.5mm} 0$, $\forall k$, the estimation variances of $\tau_{\mathrm{e}}[n]$, $\nu_{\mathrm{e}}[n]$, $\phi_{\mathrm{e}}[n]$, and $\theta_{\mathrm{e}}[n]$ can be modeled by $\sigma^2_{\tau _{\rm{e}}[n]} \hspace{-1mm}=\hspace{-1mm} c_{{\tau _{\rm{e}}}}/\mathrm{SNR}$, $\sigma^2_{\nu _{\rm{e}}[n]} \hspace{-1mm}=\hspace{-1mm} c_{{\nu _{\rm{e}}}}/\mathrm{SNR}$, $\sigma^2_{\theta_{\mathrm{e}}[n]} \hspace{-1mm}= \hspace{-1mm}c_{{\theta _{\rm{e}}}}/\mathrm{SNR}$, and $\sigma^2_{ \phi_{\mathrm{e}}[n]} \hspace{-1mm}=\hspace{-1mm} c_{{\phi _{\rm{e}}}}/\mathrm{SNR}$, respectively, where the MF output signal-to-noise ratio (SNR) is given by
\vspace{-3mm} 
\begin{equation}\label{MFsnr}
	\mathrm{SNR} \hspace{-1mm}=\hspace{-1mm} \frac{\vartheta_{\mathrm{e}}\beta^2_0G_{\mathrm{MF}}M^{\mathrm{r}}_{\mathrm{b}} \big|\mathbf{A}^{\mathrm{H}}_{{M^{\mathrm{tx}}_{\mathrm{b}}},{M^{\mathrm{ty}}_{\mathrm{b}}}}\left(\theta_{\mathrm{e}}[n], \phi_{\mathrm{e}}[n]\right) \mathbf{w}_{\mathrm{e}} [n]\big|^2}{16\pi \sigma_{\mathrm{b}}^2d^4_{\mathrm{e}}[n]}.\vspace{-3mm}
\end{equation}
Parameters $c_{{\tau _{\rm{e}}}}, c_{\nu _{\rm{e}}}, c_{\theta_{\mathrm{be}}}, c_{\phi_{\mathrm{be}}}\hspace{-1mm}>\hspace{-1mm}0$ are determined by the specific adopted estimation methods \cite{FanLiuISAC} and $G_{\mathrm{MF}}$ is the MF gain, which is proportional to the number of transmit symbols in a time slot.

\vspace{-4mm}
\subsection{Measurement Model for E-UAV}
\vspace{-3mm}
The measurement model characterizes the relationship between the observable parameters and the hidden state, which is the key for inferring the state of E-UAV.
%
%
In particular, considering the positions of both I-UAV and E-UAV in Fig. \ref{fig:SystemModel}, the measurement models associated with $\tau_{\mathrm{e}}\hspace{-0.25mm}[\hspace{-0.25mm}n\hspace{-0.25mm}]$, $\nu_{\mathrm{e}}\hspace{-0.25mm}[\hspace{-0.25mm}n\hspace{-0.25mm}]$, $\phi_{\mathrm{e}}\hspace{-0.25mm}[\hspace{-0.25mm}n\hspace{-0.25mm}]$, and $\theta_{\mathrm{e}}\hspace{-0.25mm}[\hspace{-0.25mm}n\hspace{-0.25mm}]$ are given by
\vspace{-3.25mm}
\begin{align}\label{MeasurementModel}
\hat{\tau} _{\rm{e}}[n] &\hspace{-1mm}=\hspace{-1mm} \frac{2\|\mathbf{q}_{\mathrm{e}} [n] - \mathbf{q}_{{\mathrm{b}}}\|}{c} + v_{\tau_{{\mathrm{e}}}[n]},\notag\\[-1mm]
{\hat{\nu} _{\rm{e}}}[n] &\hspace{-1mm}=\hspace{-1mm} \frac{2\dot{\mathbf{q}}^{\mathrm{T}}_{\mathrm{e}} [n] \left(\mathbf{q}_{\mathrm{e}} [n] \hspace{-1mm}-\hspace{-1mm} \mathbf{q}_{{\mathrm{b}}}\right) f_c}{c\|\mathbf{q}_{\mathrm{e}} [n] - \mathbf{q}_{{\mathrm{b}}}\|} + v_{\nu _{\rm{e}}[n]},\notag\\[-1mm]
\sin \hat{\theta}_{\mathrm{e}}[n] &\hspace{-1mm}=\hspace{-1mm} \frac{x_{\mathrm{e}}[n] \hspace{-1mm}-\hspace{-1mm} x_{\mathrm{b}}}{\sqrt{\left|x_{\mathrm{e}}[n] \hspace{-1mm}-\hspace{-1mm} x_{\mathrm{b}}\right|^2\hspace{-1mm}+\hspace{-1mm}\left|y_{\mathrm{e}}[n] \hspace{-1mm}-\hspace{-1mm} y_{\mathrm{b}}\right|^2}} \hspace{-1mm}+\hspace{-1mm} v_{\sin \theta_{\mathrm{e}}[n]},\notag\\[-1mm]
\cos \hat{\theta}_{\mathrm{e}}[n] &\hspace{-1mm}=\hspace{-1mm} \frac{y_{\mathrm{e}}[n] \hspace{-1mm}-\hspace{-1mm} y_{\mathrm{b}}}{\sqrt{\left|x_{\mathrm{e}}[n] \hspace{-1mm}-\hspace{-1mm} x_{\mathrm{b}}\right|^2\hspace{-1mm}+\hspace{-1mm}\left|y_{\mathrm{e}}[n] \hspace{-1mm}-\hspace{-1mm} y_{\mathrm{b}}\right|^2}} \hspace{-1mm}+\hspace{-1mm} v_{\cos \theta_{\mathrm{e}}[n]},\;\text{and}\notag\\[-1mm]
\sin \hat{\phi}_{\mathrm{e}}[n] &\hspace{-1mm}=\hspace{-1mm} \frac{z_{\mathrm{e}}[n] - z_{\mathrm{b}}}{\|\mathbf{q}_{\mathrm{e}} [n] - \mathbf{q}_{{\mathrm{b}}}\|} + v_{\sin \phi_{\mathrm{e}}[n]},
\end{align}
\vspace{-5.5mm}\par\noindent
respectively, where $f_c$ is the carrier frequency and $c$ is the light speed. 
$v_{\tau_{{\mathrm{e}}}[\hspace{-0.25mm}n\hspace{-0.25mm}]}$, $v_{\nu _{\rm{e}}[\hspace{-0.25mm}n\hspace{-0.25mm}]}$, $v_{\sin \theta_{\mathrm{e}}[\hspace{-0.25mm}n\hspace{-0.25mm}]}$, $v_{\cos \theta_{\mathrm{e}}[\hspace{-0.25mm}n\hspace{-0.25mm}]}$, and $v_{\sin \phi_{\mathrm{e}}[\hspace{-0.25mm}n\hspace{-0.25mm}]}$ denote measurement noises with variances $\sigma^2_{\tau _{\rm{e}}[\hspace{-0.25mm}n\hspace{-0.25mm}]}$, $\sigma^2_{\nu _{\rm{e}}[\hspace{-0.25mm}n\hspace{-0.25mm}]}$, $\sigma^2_{\sin \theta_{\mathrm{e}}[\hspace{-0.25mm}n\hspace{-0.25mm}]}$, $\sigma^2_{\cos \theta_{\mathrm{e}}[\hspace{-0.25mm}n\hspace{-0.25mm}]}$, and $\sigma^2_{\sin \phi_{\mathrm{e}}[\hspace{-0.25mm}n\hspace{-0.25mm}]}$, respectively.
Using trigonometric identities and assuming $\sigma^2_{\theta_{\mathrm{e}}[\hspace{-0.25mm}n\hspace{-0.25mm}]},\sigma^2_{\phi_{\mathrm{e}}[\hspace{-0.25mm}n\hspace{-0.25mm}]}\hspace{-1mm}\to\hspace{-1mm} 0$, we have $\sigma_{\sin \theta_{\mathrm{e}}[\hspace{-0.25mm}n\hspace{-0.25mm}]} \hspace{-1mm}\approx\hspace{-1mm} \cos \hat{\theta}_{\mathrm{e}}[\hspace{-0.25mm}n\hspace{-0.25mm}] \sigma_{\theta_{\mathrm{e}}[\hspace{-0.25mm}n\hspace{-0.25mm}]}$, $\sigma_{\cos \theta_{\mathrm{e}}[\hspace{-0.25mm}n\hspace{-0.25mm}]} \hspace{-1mm}\approx\hspace{-1mm} \sin \hat{\theta}_{\mathrm{e}}[\hspace{-0.25mm}n\hspace{-0.25mm}] \sigma_{\theta_{\mathrm{e}}[\hspace{-0.25mm}n\hspace{-0.25mm}]}$, and $\sigma_{\sin \phi_{\mathrm{e}}[\hspace{-0.25mm}n\hspace{-0.25mm}]} \hspace{-1mm}\approx\hspace{-1mm} \cos \hat{\phi}_{\mathrm{e}}[\hspace{-0.25mm}n\hspace{-0.25mm}] \sigma_{\phi_{\mathrm{e}}[\hspace{-0.25mm}n\hspace{-0.25mm}]}$.

Let $\boldsymbol{\beta}_{\mathrm{e}} [\hspace{-0.25mm}n\hspace{-0.25mm}] \hspace{-1mm}= \hspace{-1mm}[\hat{\tau} _{\rm{e}}[\hspace{-0.25mm}n\hspace{-0.25mm}],{\hat{\nu} _{\rm{e}}}[\hspace{-0.25mm}n\hspace{-0.25mm}], \sin \hat{\theta}_{\mathrm{e}}[\hspace{-0.25mm}n\hspace{-0.25mm}], \cos \hat{\theta}_{\mathrm{e}}[\hspace{-0.25mm}n\hspace{-0.25mm}], \sin \hat{\phi}_{\mathrm{e}}[\hspace{-0.25mm}n\hspace{-0.25mm}]]^\mathrm{T} \hspace{-1mm}\in\hspace{-1mm} \mathbb{R}^{5 \times 1}$ collect all observable parameters. Then, the measurement model can be rewritten as
\vspace{-3.5mm}
\begin{equation}\label{MeautrementModel2}
\boldsymbol{\beta}_{\mathrm{e}} [n] =  \mathbf{g}_n\left(\boldsymbol{\alpha}_{\mathrm{e}} [n]\right) + \mathbf{v}_{\boldsymbol{\beta}_{\mathrm{e}}[n]},\vspace{-3.5mm}
\end{equation}
where $\mathbf{v}_{\boldsymbol{\beta}_{\mathrm{e}}[n]}\hspace{-0.5mm} \hspace{-1mm}\sim\hspace{-1mm} \mathcal{N}\hspace{-1mm}\left(\hspace{-0.25mm}\mathbf{0},\hspace{-0.25mm}\mathbf{Q}_{\boldsymbol{\beta}_{\mathrm{e}}[\hspace{-0.25mm}n\hspace{-0.25mm}]}\hspace{-0.25mm}\right)$ and $\mathcal{N}\hspace{-1mm}\left(\hspace{-0.25mm}\boldsymbol{\mu},\hspace{-0.25mm}\boldsymbol{\Sigma}\hspace{-0.25mm}\right)$ represents the Gaussian distribution with mean $\boldsymbol{\mu}$ and covariance matrix $\boldsymbol{\Sigma}$.
The main diagonal entries of the measurement covariance matrix $\mathbf{Q}_{\boldsymbol{\beta}_{\mathrm{e}}[\hspace{-0.25mm}n\hspace{-0.25mm}]}$ are given by $\sigma^2_{\tau _{\rm{e}}\hspace{-0.25mm}[\hspace{-0.25mm}n\hspace{-0.25mm}]}$, $\sigma^2_{\nu _{\rm{e}}\hspace{-0.25mm}[\hspace{-0.25mm}n\hspace{-0.25mm}]}$, $\sigma^2_{\sin \hspace{-0.25mm} \theta_{\mathrm{e}}\hspace{-0.25mm}[\hspace{-0.25mm}n\hspace{-0.25mm}]}$, $\sigma^2_{\cos \hspace{-0.25mm} \theta_{\mathrm{e}}\hspace{-0.25mm}[\hspace{-0.25mm}n\hspace{-0.25mm}]}$, and $\sigma^2_{\sin \hspace{-0.25mm} \phi_{\mathrm{e}}\hspace{-0.25mm}[\hspace{-0.25mm}n\hspace{-0.25mm}]}$, respectively, and the only two non-zero off-diagonal entries are $\left\{\hspace{-0.5mm}\mathbf{Q}_{\boldsymbol{\beta}_{\mathrm{e}}\hspace{-0.25mm}[\hspace{-0.25mm}n\hspace{-0.25mm}]}\hspace{-0.5mm}\right\}_{3,4} \hspace{-1mm}=\hspace{-1mm} \left\{\hspace{-0.5mm}\mathbf{Q}_{\boldsymbol{\beta}_{\mathrm{e}}\hspace{-0.25mm}[\hspace{-0.25mm}n\hspace{-0.25mm}]}\hspace{-0.5mm}\right\}_{4,3} \hspace{-1mm}= \hspace{-1mm}\sigma_{\cos\hspace{-0.25mm} \theta_{\mathrm{e}}\hspace{-0.25mm}[\hspace{-0.25mm}n\hspace{-0.25mm}]}\sigma_{\sin\hspace{-0.25mm} \theta_{\mathrm{e}}\hspace{-0.25mm}[\hspace{-0.25mm}n\hspace{-0.25mm}]}$.
The non-linear function $\mathbf{g}_n\hspace{-1mm}:\hspace{-1mm} \mathbb{R}^{6 \times 1} \hspace{-1mm}\to\hspace{-1mm} \mathbb{R}^{5 \times 1}$ represents the measurement functions defined by \eqref{MeasurementModel}.
%
%
%

%

\vspace{-4mm}
\subsection{State Evolution Model of E-UAV}
\vspace{-3mm}
%
Assuming a constant velocity movement model \cite{FanLiuISAC}, the state evolution model of E-UAV is given by
\vspace{-3.5mm}
\begin{equation}\label{StateEvolution}
\boldsymbol{\alpha}_{\mathrm{e}} [n] =  \mathbf{F}\boldsymbol{\alpha}_{\mathrm{e}} [n-1] + \mathbf{v}_{\boldsymbol{\alpha}_{\mathrm{e}}}[n],\vspace{-3.5mm}
\end{equation}
where $\mathbf{F}\hspace{-1mm}\in \hspace{-1mm}\mathbb{R}^{6 \times 6}$ is the state transition matrix.
All main diagonal entries of $\mathbf{F}$ are one and the only three non-zero off-diagonal entries are $\left\{\hspace{-0.5mm}\mathbf{F}\hspace{-0.5mm}\right\}_{1,4} \hspace{-1mm}=\hspace{-1mm} \left\{\hspace{-0.5mm}\mathbf{F}\hspace{-0.5mm}\right\}_{2,5} \hspace{-1mm}=\hspace{-1mm} \left\{\hspace{-0.5mm}\mathbf{F}\hspace{-0.5mm}\right\}_{3,6} \hspace{-1mm}=\hspace{-1mm} \delta$.
Vector $\mathbf{v}_{\boldsymbol{\alpha}_{\mathrm{e}}[\hspace{-0.25mm}n\hspace{-0.25mm}]} \hspace{-1mm}\sim \hspace{-1mm} \mathcal{N}\hspace{-1mm}\left(\hspace{-0.25mm}\mathbf{0},\mathbf{Q}_{\boldsymbol{\alpha}_{\mathrm{e}}}\hspace{-0.25mm}\right)$ is the state evolution noise vector and $\mathbf{Q}_{\boldsymbol{\alpha}_{\mathrm{e}}}$ is the state evolution covariance matrix, which is assumed to be known by I-UAV from long-term measurements. 

\vspace{-4mm}
\subsection{Tracking E-UAV via EKF}
\vspace{-3mm}
Due to the non-linearity of the measurement model in \eqref{MeautrementModel2}, we adopt an extended Kalman filter (EKF) \cite{anderson2012optimal} to track E-UAV.
In time slot $n$, we assume that I-UAV has the estimates of E-UAV's state in the previous time slot, $\hat{\boldsymbol{\alpha}}_{\mathrm{e}} [n\hspace{-1mm}-\hspace{-1mm}1]$, with the corresponding covariance matrix $\mathbf{C}_{\mathrm{e}}[n\hspace{-1mm}-\hspace{-1mm}1] \hspace{-1mm} \in \hspace{-1mm} \mathbb{R}^{6 \times 6}$.
I-UAV predicts the state of E-UAV in time slot $n$ by $\hat{\boldsymbol{\alpha}}_{\mathrm{e}} [n|n\hspace{-1mm}-\hspace{-1mm}1] \hspace{-1mm}=\hspace{-1mm} \mathbf{F}\hat{\boldsymbol{\alpha}}_{\mathrm{e}} [n\hspace{-1mm}-\hspace{-1mm}1]$, where the prediction covariance matrix is given by
\vspace{-3.5mm}
\begin{equation}\label{CovarianceMatrixPrediction}
\mathbf{C}_{\mathrm{e}}[n|n-1] = \mathbf{F}\mathbf{C}_{\mathrm{e}}[n-1]\mathbf{F}^{\mathrm{T}} + \mathbf{Q}_{\boldsymbol{\alpha}_{\mathrm{e}}}.\vspace{-3.5mm}
\end{equation}
According to the predicted state, one can predict the effective channel between I-UAV and E-UAV in time slot $n$ as follows,
\vspace{-3.5mm}
\begin{equation}\label{eqn:PredictedChannel}
\hat{\mathbf{h}}_{\mathrm{be}}[n|n\hspace{-1mm}-\hspace{-1mm}1] \hspace{-1mm}=\hspace{-1mm} \frac{\beta_0}{\hat{d}_{\mathrm{e}}\hspace{-0.25mm}[\hspace{-0.25mm}n|n\hspace{-1mm}-\hspace{-1mm}1\hspace{-0.25mm}]}\mathbf{A}_{{M^{\mathrm{tx}}_{\mathrm{b}}},{M^{\mathrm{ty}}_{\mathrm{b}}}}\hspace{-1mm}\left(\hspace{-0.5mm}\hat{\theta}_{\mathrm{e}}\hspace{-0.25mm}[\hspace{-0.25mm}n|n\hspace{-1mm}-\hspace{-1mm}1\hspace{-0.25mm}], \hat{\phi}_{\mathrm{e}}\hspace{-0.25mm}[\hspace{-0.25mm}n|n\hspace{-1mm}-\hspace{-1mm}1\hspace{-0.25mm}]\hspace{-0.5mm}\right),\vspace{-3.5mm}
\end{equation}
where $\hat{d}_{\mathrm{e}}[n|n\hspace{-1mm}-\hspace{-1mm}1]$, $\hat{\theta}_{\mathrm{e}}[n|n\hspace{-1mm}-\hspace{-1mm}1]$, and $\hat{\phi}_{\mathrm{e}}[n|n\hspace{-1mm}-\hspace{-1mm}1]$ are the predicted distance, azimuth AOD, and elevation AOD, respectively.
Due to the prediction uncertainty in \eqref{CovarianceMatrixPrediction}, the channel prediction in \eqref{eqn:PredictedChannel} is also inaccurate.
However, analyzing the prediction variance of ${\mathbf{h}}_{\mathrm{be}}[n]$ is a challenging task due to complicated mapping function between ${\boldsymbol{\alpha}}_{\mathrm{e}} [n]$ and ${\mathbf{h}}_{\mathrm{be}}[n]$.
To facilitate robust resource allocation design, we assume $\left\|\Delta {\mathbf{h}}_{\mathrm{be}}[n|n\hspace{-1mm}-\hspace{-1mm}1]\right\|^2 \hspace{-1mm}\le\hspace{-1mm} \sigma^2_{{\mathbf{h}}_{\mathrm{be}}[n|n\hspace{-0.25mm}-\hspace{-0.25mm}1]}$ holds for the channel prediction error $\Delta {\mathbf{h}}_{\mathrm{be}}[n|n\hspace{-1mm}-\hspace{-1mm}1]$, where $\sigma^2_{{\mathbf{h}}_{\mathrm{be}}[n|n\hspace{-0.25mm}-\hspace{-0.25mm}1]} \hspace{-1mm}=\hspace{-1mm} \left\|\boldsymbol{\gamma}^{\mathrm{H}}\mathbf{C}_{\mathrm{e}}[n|n\hspace{-1mm}-\hspace{-1mm}1]\right\|^2$ and parameters $\boldsymbol{\gamma} \in \mathbb{R}^{6 \times 1}$ can be obtained via Monte Carlo simulation \cite{fishman2013monte}.

%

To update the state of E-UAV in time slot $n$, we linearize the measurement model in \eqref{MeautrementModel2} around the predicted state, i.e.,
\vspace{-3.5mm}
\begin{equation}\label{MeautrementModelLinearization}
\boldsymbol{\beta}_{\mathrm{e}} \hspace{-0.25mm}[\hspace{-0.25mm}n\hspace{-0.25mm}] \hspace{-0.5mm}\approx\hspace{-0.5mm} \mathbf{g}_n\hspace{-1mm}\left(\hat{\boldsymbol{\alpha}}_{\mathrm{e}} [n|n\hspace{-1mm}-\hspace{-1mm}1]\right)\hspace{-0.5mm} +\hspace{-0.5mm} \mathbf{G}_n\hspace{-1mm}\left({\boldsymbol{\alpha}}_{\mathrm{e}}\hspace{-0.25mm} [\hspace{-0.25mm}n\hspace{-0.25mm}] \hspace{-1mm}-\hspace{-1mm} \hat{\boldsymbol{\alpha}}_{\mathrm{e}}\hspace{-0.25mm} [n|n\hspace{-1mm}-\hspace{-1mm}1]\right)\hspace{-0.5mm}+\hspace{-0.5mm} \mathbf{v}_{\boldsymbol{\beta}_{\mathrm{e}}}\hspace{-0.5mm}[\hspace{-0.25mm}n\hspace{-0.25mm}],\hspace{-1mm}\vspace{-3.5mm}
\end{equation}
where $\mathbf{G}_n \hspace{-1mm}\in\hspace{-1mm} \mathbb{R}^{5 \hspace{-0.25mm}\times\hspace{-0.25mm} 6}$ denotes the Jacobian matrix for $\mathbf{g}_n$ with respect to (w.r.t.) ${\boldsymbol{\alpha}}_{\mathrm{e}} [\hspace{-0.25mm}n\hspace{-0.25mm}]$ for the predicted state, i.e., $\mathbf{G}_n \hspace{-1mm}= \hspace{-1mm}\frac{\partial \mathbf{g}_n }{\partial {\boldsymbol{\alpha}}_{\mathrm{e}} [\hspace{-0.25mm}n\hspace{-0.25mm}]} |_{\hat{\boldsymbol{\alpha}}_{\mathrm{e}} [\hspace{-0.25mm}n|n-1\hspace{-0.25mm}]}$.
Now, the state of E-UAV in time slot $n$ is estimated as follows,
\vspace{-3.5mm}
\begin{equation}\label{StateTracking}
\hat{{\boldsymbol{\alpha}}}_{\mathrm{e}} [n] = \hat{\boldsymbol{\alpha}}_{\mathrm{e}} [n|n-1] + \mathbf{K}_n\left(\boldsymbol{\beta}_{\mathrm{e}} [n] - \mathbf{g}_n\left(\hat{\boldsymbol{\alpha}}_{\mathrm{e}} [n|n-1]\right)\right),\vspace{-3.5mm}
\end{equation}
and the corresponding posterior covariance matrix is
\vspace{-3.5mm}
\begin{equation}\label{CovPosteriorMSE}
\mathbf{C}_{\mathrm{e}}[n] = \left(\mathbf{I}_6 - \mathbf{K}_n{{\bf{G}}_n}\right)\mathbf{C}_{\mathrm{e}}[n|n-1] \in \mathbb{R}^{6 \times 6},\vspace{-3.5mm}
\end{equation}
where the Kalman gain matrix $\mathbf{K}_n \in \mathbb{R}^{6 \times 5}$ is given by
\vspace{-3.5mm}
\begin{equation}\label{KalmanGain}
\mathbf{K}_n = \mathbf{C}_{\mathrm{e}}[n|n\hspace{-1mm}-\hspace{-1mm}1]{{\bf{G}}^{\mathrm{T}}_n}\hspace{-1mm}\left({{\bf{G}}_n}\mathbf{C}_{\mathrm{e}}[n|n\hspace{-1mm}-\hspace{-1mm}1]{{\bf{G}}^{\mathrm{T}}_n}\hspace{-1mm}+\hspace{-1mm}\mathbf{Q}_{\boldsymbol{\beta}_{\mathrm{e}}}[n]\right)^{-1}.\vspace{-3.5mm}
\end{equation}
%
%
%
Note that the trace of $\mathbf{C}_{\mathrm{e}}[n]$ characterizes the posterior mean square error (MSE) for tracking the state of E-UAV.

\vspace{-4mm}
\section{Online Resource Allocation Design}
\vspace{-4mm}
\label{sec:JointDesign}

\subsection{Problem Formulation}
\vspace{-3mm}
In time slot $n\hspace{-0.5mm}-\hspace{-0.5mm}1$, $\forall n$, the resource allocation design is formulated as the following optimization problem:
\vspace{-3mm}
\begin{align}\label{ProblemFormulation}
\underset{\mathcal{X}}{\mathrm{maximize}}\,\; & \sum\nolimits_{k=1}^{K} u_k[n]
\\[-1mm]
\notag\mbox{s.t.}\;\;
\mbox{{C1:}}\; &\sum\nolimits_{k = 1}^{K} {u}_k [n] \left\|\mathbf{w}_k [n]\right\|^2 + \left\|\mathbf{w}_{\mathrm{e}} [n]\right\|^2 \le p_{\mathrm{max}},\notag\\[-1mm]
\mbox{{C2:}}\; & {{R}_k}[n] \ge {u}_k [n] {R_{\mathrm{min}}}, \forall k,\notag\\[-1mm]
\mbox{{C3:}}\; & 
\underset{\Delta{\bf{h}}_{{\rm{be}}}[n|n-1]}{\mathrm{max}}
 {{R}^k_{\rm{e}}}[n] \le {u}_k [n] {R_{\mathrm{Leakage}}}, \forall k, \notag\\[-1mm]
\mbox{{C4:}}\; & \mathrm{Tr}\{\hat{\mathbf{C}}_{\mathrm{e}}[n]\} \le \mathrm{MSE}_{\mathrm{max}}, \notag
\end{align}
\vspace{-7.5mm}\par\noindent
where $\mathcal{X}\hspace{-1mm} =\hspace{-1mm} \{{u}_k [\hspace{-0.25mm}n\hspace{-0.25mm}] \hspace{-1mm}\in\hspace{-1mm} \{0,1\}, \mathbf{w}_k [\hspace{-0.25mm}n\hspace{-0.25mm}], \mathbf{w}_{\mathrm{e}} [\hspace{-0.25mm}n\hspace{-0.25mm}]\}$.
The objective function $\sum_{k=1}^{K} \hspace{-0.5mm}u_k[\hspace{-0.25mm}n\hspace{-0.25mm}]$ represents the number of GUs that can be served securely.
In \eqref{ProblemFormulation}, $p_{\mathrm{max}}$ in C1 is the maximum transmit power, ${R_{\mathrm{min}}}$ in C2 the minimum data rate for the selected GUs, and $\mathrm{MSE}_{\mathrm{max}}$ in C4 is the maximum tolerable tracking MSE for the location and velocity of E-UAV.
${R_{\mathrm{Leakage}}}$ in C3 is the maximum allowable leakage rate associated with the selected GUs in the presence of channel prediction uncertainty.
Note that $\mathbf{Q}_{\boldsymbol{\beta}_{\mathrm{e}}}[\hspace{-0.25mm}n\hspace{-0.25mm}]$  in \eqref{KalmanGain} depends on the MF output SNR in \eqref{MFsnr} and thus $\mathbf{C}_{\mathrm{e}}\hspace{-0.25mm}[\hspace{-0.25mm}n\hspace{-0.25mm}]$ in  \eqref{CovPosteriorMSE} depends on the values of $\theta_{\mathrm{e}}\hspace{-0.25mm}[\hspace{-0.25mm}n\hspace{-0.25mm}]$, $\phi_{\mathrm{e}}\hspace{-0.25mm}[\hspace{-0.25mm}n\hspace{-0.25mm}]$, and $d_{\mathrm{e}}\hspace{-0.25mm}[\hspace{-0.25mm}n\hspace{-0.25mm}]$, which are unknown at the time of resource allocation design.
Thus, instead of using the posterior covariance matrix $\mathbf{C}_{\mathrm{e}}\hspace{-0.25mm}[\hspace{-0.25mm}n\hspace{-0.25mm}]$, its predicted value, $\hat{\mathbf{C}}_{\mathrm{e}}\hspace{-0.25mm}[\hspace{-0.25mm}n\hspace{-0.25mm}]$, is adopted in C4.
Substituting \eqref{KalmanGain} into \eqref{CovPosteriorMSE} and employing the matrix inversion lemma, $\hat{\mathbf{C}}^{-1}_{\mathrm{e}}\hspace{-0.25mm}[\hspace{-0.25mm}n\hspace{-0.25mm}]$ is given by
\vspace{-3mm}
\begin{equation}\label{CovPosteriorMSEInverse}
\hat{\mathbf{C}}^{-1}_{\mathrm{e}}[n] = \mathbf{C}^{-1}_{\mathrm{e}}[n|n-1] + {{\bf{G}}^{\mathrm{T}}_n}\hat{\mathbf{Q}}^{-1}_{\boldsymbol{\beta}_{\mathrm{e}}[n]}{{\bf{G}}_n},\vspace{-4mm}
\end{equation}
where $\hat{\mathbf{Q}}_{\boldsymbol{\beta}_{\mathrm{e}}\hspace{-0.25mm}[\hspace{-0.25mm}n\hspace{-0.25mm}]}$ is the predicted measurement covariance matrix, which can be obtained based on the predicted state of the E-UAV $\hat{\boldsymbol{\alpha}}_{\mathrm{e}} [\hspace{-0.25mm}n|n\hspace{-1mm}-\hspace{-1mm}1\hspace{-0.25mm}]$.
Note that the prediction uncertainty is taken into account in C4 via $\mathbf{C}_{\mathrm{e}}[\hspace{-0.25mm}n|n\hspace{-1mm}-\hspace{-1mm}1\hspace{-0.25mm}]$, see \eqref{CovPosteriorMSEInverse}.

\vspace{-5mm}
\subsection{Proposed Solution}
\vspace{-3mm}
\label{subsec:ProposedSolution}
Now, introducing six auxiliary variables $t_i \hspace{-0.5mm} \ge \hspace{-0.5mm} 0$, $i \hspace{-0.5mm}=\hspace{-0.5mm} \{1,\ldots,6\}$, to bound the main diagonal entries of $\hat{\mathbf{C}}_{\mathrm{e}}\hspace{-0.25mm}[\hspace{-0.25mm}n\hspace{-0.25mm}]$, i.e., $\{\hat{\mathbf{C}}_{\mathrm{e}}[\hspace{-0.25mm}n\hspace{-0.25mm}]\}_{ii} \hspace{-1mm}\le\hspace{-1mm} t_i$, C4 in \eqref{ProblemFormulation} becomes $\mbox{{C4:}} \sum\nolimits_{i=1}^{6} \hspace{-0.5mm}t_i\hspace{-1mm} \le\hspace{-1mm} \mathrm{MSE}_{\mathrm{max}}$.
The problem in \eqref{ProblemFormulation} is then equivalently transformed as follows
\vspace{-4mm}
\begin{equation}\label{ProblemFormulationII}
\underset{\mathcal{X},t_i}{\mathrm{minimize}} 
\sum\nolimits_{k=1}^{K} \hspace{-1mm} u_k[\hspace{-0.25mm}n\hspace{-0.25mm}]\;\;
\mbox{s.t.}\;
\mbox{{C1-C4}},\mbox{{C5:}}\left[\hspace{-2mm} {\begin{array}{*{20}{c}}
	{\hat{\bf{C}}_{\rm{e}}^{ - 1}[n]}&\hspace{-2mm}{{{\bf{e}}_i}}\\
	{{\bf{e}}_i^{\rm{T}}}&\hspace{-2mm}{{t_i}}
	\end{array}} \hspace{-2mm}\right]\hspace{-1mm} \succeq \hspace{-1mm}\mathbf{0},\vspace{-4mm}
\end{equation}
where ${{{\bf{e}}_i}} \hspace{-1mm}\in\hspace{-1mm} \mathbb{R}^{6 \times 1}$ is the $i$-th column of the $6 \hspace{-1mm}\times\hspace{-1mm} 6$ identity matrix.

One obstacle for solving \eqref{ProblemFormulationII} is the binary variable ${u}_k [n]$.
Conventional approaches to handle binary variables, such as the big-M formulation and successive convex approximation \cite{WeiNOMA7934461}, require iterative algorithms, whose computational complexity and delay might not be affordable for online resource allocation design.
Fortunately, for given user scheduling variables, the problem in \eqref{ProblemFormulationII} becomes a feasibility problem which
can be solved optimally by the commonly-used \textit{S-procedure} \cite{book:convex} and semidefinite relaxation (SDR) approach \cite{ZhiQuanTSPM}.
Hence, we propose a channel correlation-based user scheduling strategy.
Without loss of generality, all GUs are indexed in descending order of the correlation coefficients between $\hat{\mathbf{h}}_{\mathrm{be}}[n|n\hspace{-1mm}-\hspace{-1mm}1]$ and $\mathbf{h}_{\mathrm{b},k}[n]$, $\forall k$.
In general, the higher the channel correlation, the higher the risk of information leakage. 
Then, we first select all GUs for service, i.e., $u_k[n] = 1$, $\forall k$.
If the resulting problem in \eqref{ProblemFormulationII} is infeasible, we de-select GUs one-by-one in descending order of their channel correlations until \eqref{ProblemFormulationII} becomes feasible.
If all GUs are de-selected and the problem in \eqref{ProblemFormulationII} is still infeasible, the I-UAV transmits only AN for jamming and sensing adopting maximum ratio transmission, i.e., $\mathbf{w}_k [n] = \mathbf{0}$, $\forall k$, and $\mathbf{w}_{\mathrm{e}} [n] = \frac{\hat{\mathbf{h}}_{\mathrm{be}}[n|n\hspace{-0.25mm}-\hspace{-0.25mm}1]}{\|\hat{\mathbf{h}}_{\mathrm{be}}[n|n\hspace{-0.25mm}-\hspace{-0.25mm}1]\|}\sqrt{p_{\mathrm{max}}}$.
The details of the resulting algorithm are omitted here due to space limitation.

\begin{figure}[t]
	\vspace{-2mm}
	\center{\includegraphics[width=2.8in]{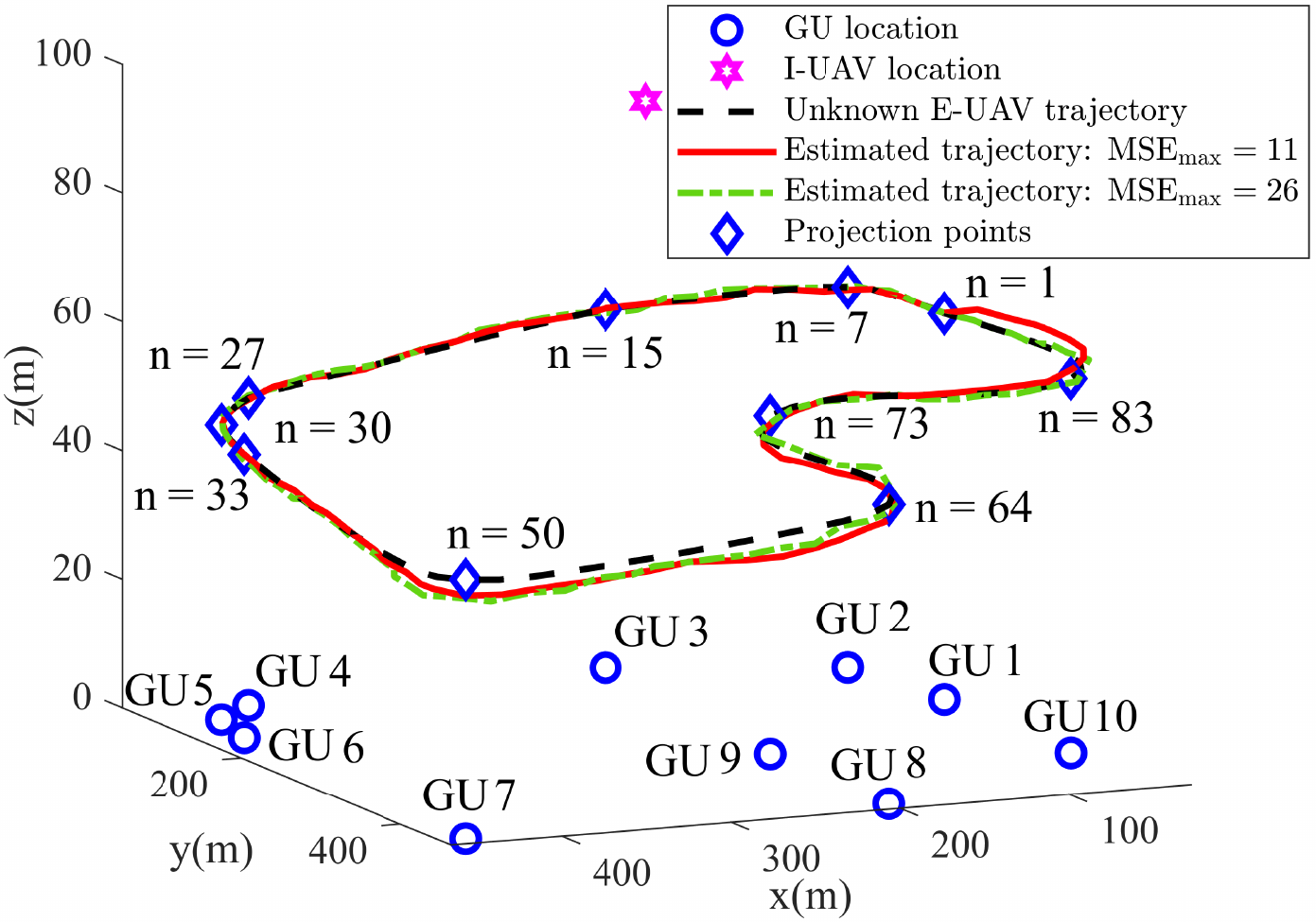}}\vspace{-4mm}
	\caption{Estimated trajectory of E-UAV.}\vspace{-7mm}
	\label{fig:ISJCResult}
\end{figure}

\vspace{-4mm}
\section{Simulation Results}
\vspace{-4mm}
\label{sec:SimResults}
In this section, we evaluate the performance of the proposed ISJC scheme via simulations.
The main simulation parameters are as follows: $K\hspace{-1mm}=\hspace{-1mm}10$, $\delta\hspace{-1mm}=\hspace{-1mm}1$ s, $N = 92$,  $p_{\mathrm{max}}\hspace{-1mm}=\hspace{-1mm}30$ dBm, ${R_{\mathrm{min}}}\hspace{-1mm}=\hspace{-1mm}5$ bit/s/Hz, ${R_{\mathrm{Leakage}}}\hspace{-1mm}=\hspace{-1mm}0.01$ bit/s/Hz,
$G_{\mathrm{MF}} = 10^4$, $\sigma^2_{\rm{e}}\hspace{-1mm}=\hspace{-1mm}\sigma^2_k\hspace{-1mm}=\hspace{-1mm}-99$ dBm, $\sigma^2_{\rm{b}}=-50$ dBm, $\beta^2_0\hspace{-1mm}=\hspace{-1mm}50$ dB, $\vartheta_{\mathrm{e}}\hspace{-1mm}=\hspace{-1mm}0.1$ $\rm m^2$, and ${M^{\mathrm{tx}}_{\mathrm{b}}}\hspace{-1mm}=\hspace{-1mm}{M^{\mathrm{ty}}_{\mathrm{b}}}\hspace{-1mm}=\hspace{-1mm}{M^{\mathrm{rx}}_{\mathrm{b}}}\hspace{-1mm}=\hspace{-1mm}{M^{\mathrm{ry}}_{\mathrm{b}}}\hspace{-1mm}=\hspace{-1mm}{M^{\mathrm{rx}}_{\mathrm{e}}}\hspace{-1mm}=\hspace{-1mm}{M^{\mathrm{ry}}_{\mathrm{e}}}\hspace{-1mm}=\hspace{-1mm}4$.
Note that a higher $\mathrm{MSE}_{\mathrm{max}}$ implies that a poorer tracking performance is tolerable by the considered system.
Thus, we consider $\mathrm{MSE}_{\mathrm{max}} \hspace{-1mm}\in \hspace{-1mm}\left[5,26\right]$.
The locations of I-UAV and the $K$ GUs and the unknown trajectory of E-UAV are illustrated in Fig. \ref{fig:ISJCResult}.
The modeling parameters, including $c_{{\tau _{\rm{e}}}}$, $c_{\nu _{\rm{e}}}$, $c_{\theta_{\mathrm{be}}}$, $c_{\phi_{\mathrm{be}}}$, and $\mathbf{Q}_{\boldsymbol{\alpha}_{\mathrm{e}}}$, are initialized numerically assuming {random walks} of the I-UAV and E-UAV, respectively.

\begin{figure}[t]
	\vspace{-2mm}
	\center{\includegraphics[width=2.9in]{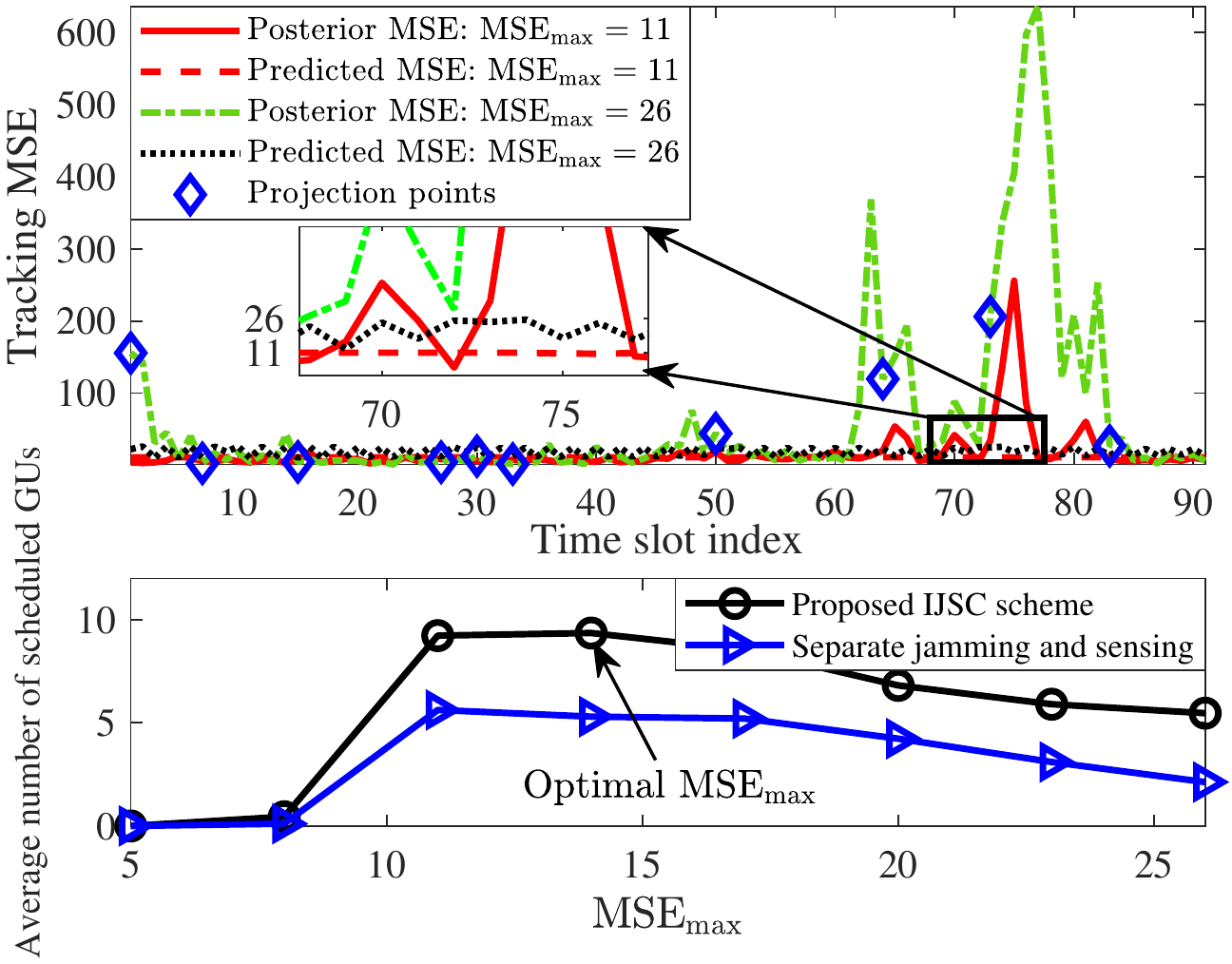}}\vspace{-3mm}
	\caption{Tracking (upper half) and secrecy communications (lower half) performance.}\vspace{-7mm}
	\label{fig:TrackingPerformance}
\end{figure}

The estimated trajectory of E-UAV is shown in Fig. \ref{fig:ISJCResult}.
The corresponding posterior tracking MSE, $\left\|\hat{{\boldsymbol{\alpha}}}_{\mathrm{e}} [\hspace{-0.25mm}n\hspace{-0.25mm}] \hspace{-1mm}-\hspace{-1mm} {{\boldsymbol{\alpha}}}_{\mathrm{e}} [\hspace{-0.25mm}n\hspace{-0.25mm}]\right\|^2$, in each time slot is shown in the upper half of Fig. \ref{fig:TrackingPerformance}.
We observe that a smaller $\mathrm{MSE}_{\mathrm{max}}$ leads to a lower tracking MSE and a more accurate trajectory estimation.
In fact, a smaller $\mathrm{MSE}_{\mathrm{max}}$ imposes a more stringent tracking MSE constraint for resource allocation design and thus more power and spatial degrees of freedom are allocated for sensing.
%
%
To provide more insights, we define $K$ ``projection points'' on the unknown trajectory for which E-UAV has the smallest 3D distance w.r.t. the GUs, respectively.
Comparing Fig. \ref{fig:ISJCResult} and Fig. \ref{fig:TrackingPerformance}, we observe a high tracking MSE around some projection points, such as $n \hspace{-1mm}=\hspace{-1mm} 64$ and $n \hspace{-1mm}= \hspace{-1mm}73$, where E-UAV changes directions, since the constant velocity model assumption for the EKF is less accurate in those points.
Note that in Fig. \ref{fig:TrackingPerformance}, the posterior tracking MSE in some time slots is higher than $\mathrm{MSE}_{\mathrm{max}}$ while the predicted MSE is guaranteed to be smaller than $\mathrm{MSE}_{\mathrm{max}}$.
In fact, $\mathrm{MSE}_{\mathrm{max}}$ only limits the predicted tracking MSE for resource allocation design while the actual tracking MSE, which is not accessible to the I-UAV at the time of resource allocation design, might be much higher due to the state evolution model mismatch and the measurement uncertainty in \eqref{MeautrementModel2}.
%
%
%
Furthermore, the average number of scheduled GUs versus $\mathrm{MSE}_{\mathrm{max}}$ is shown in the lower half of Fig. \ref{fig:TrackingPerformance}.
The performance of a baseline separate jamming and sensing scheme, where a dedicated sub-slot is used for sensing and the remaining time is used for communications and jamming, is also illustrated for comparison.
We can observe a higher number of scheduled GUs for the proposed scheme compared to the benchmark scheme.
This is because the dual use of AN in the proposed scheme preserves the system resources and provides more flexibility for resource allocation design.
This underlines the advantage of integrating sensing, jamming, and communications for improving secrecy performance.
It can be further observed that both a small $\mathrm{MSE}_{\mathrm{max}}$ and a large $\mathrm{MSE}_{\mathrm{max}}$ result in a small number of securely served GUs for both schemes.
For a small $\mathrm{MSE}_{\mathrm{max}}$, more resources are needed for sensing, and thus, fewer GUs can be scheduled. 
This implies that enhancing the tracking performance is not always beneficial for communications as it requires more system resources.
Increasing $\mathrm{MSE}_{\mathrm{max}}$ relaxes the tracking performance constraint C4 in \eqref{ProblemFormulation} and thus results in a larger objective value.
However, a too large $\mathrm{MSE}_{\mathrm{max}}$ in a given time slot leads to unreliable eavesdropper channel prediction in the next time slot, and thus, a lower jamming efficiency.
Hence, the number of securely served GUs is ultimately limited by the channel prediction uncertainty in the large $\mathrm{MSE}_{\mathrm{max}}$ regime.
In fact, choosing $\mathrm{MSE}_{\mathrm{max}}$ properly is critical for maximizing the secrecy communication performance.
For example, for the considered scenario, $\mathrm{MSE}_{\mathrm{max}} = 14$ is optimal for the proposed scheme.

\vspace{-5mm}
\section{Conclusions}
\label{sec:Conclusions}
\vspace{-4mm}

In this paper, we proposed a novel ISJC framework and an online resource allocation design for securing UAV-enabled downlink communications.
The dual use of AN enables the I-UAV to concurrently jam the E-UAV and sense the corresponding channels efficiently.
Through simulations, we demonstrated that integrating sensing, jamming, and communications improves the secrecy performance, while choosing a suitable value for the tolerable tracking MSE is critical for balancing tracking and secrecy performance.

\vfill\pagebreak

\bibliographystyle{IEEEbib}
\bibliography{UAV_ISAV}

\end{document}